\documentclass[preprint,proceedings]{rmaa}

\newcommand{\D}{\discretionary{}{}{}}

\SetYear{2003} \SetConfTitle{Winds, Bubbles and Explosions}

\title{Late helium flash in V605 Aql:\\
PNe evolution and shell formation in quick motion}

\author{
  S. Kimeswenger\altaffilmark{1}}

\altaffiltext{1}{Institut f{\"u}r Astrophysik der
Leopold--Franzens--Universit{\"a}t Innsbruck,
  Technikerstr. 25, A-6020 Innsbruck, Austria (Stefan.Kimeswenger\D{}@uibk.\D{}ac.\D{}at).}

\suppressfulladdresses

\listofauthors{S. Kimeswenger}
\indexauthor{Kimeswenger, S.}

\addkeyword{Stars: individual: V605 Aql}

\begin{document}
\maketitle

\boldabstract{I present recent observations of the born--again
core V605 Aql of the old planetary nebula Abell 58.}

The "born-again" PNe V605 Aql (A58) and V4344 Sgr (Sakurai) give
us the rare chance to follow shell formation and building of the
winds in "real-time". The high carbon abundance, and thus the high
dust formation rate, enhances the mechanisms. Thus the timescales
are shortened. My data presented here gives a  quick look on the
results of new observations of V605 Aql and its old planetary
nebula (PN) A58 obtained at the ESO NTT (August 2002).

Remarkable changes of the spectrum relative to that obtained by
Guerrero \& Manchado (1996) in June 1994 were found. The
[\ion{S}{II}]$_{6716+6731}$ and [\ion{O}{I}]$_{6300+6364}$ lines
increased compared to the [\ion{N}{II}] lines.
\ion{He}{I}$_{5875}$ raised by 50\% relative to
\ion{N}{II}$_{5754}$. Generally all lines got stronger.

The finding of Guerrero \& Manchado (1996) of an H$\alpha$ line as
strong as 30\%  of [\ion{N}{II}]$_{6548}$ is not supported by my
observations. It may originate from an inaccurate subtraction of
the line from the old nebula. The absence of H$\alpha$ was found
also by Pollacco et al. (1992). I also obtained H$\beta$ images
and the corresponding offband, obtained to remove stellar
components and continuum radiation. Those images show  the same
flux of V605 AQL in the hydrogen line and of the adjacent
continuum. The flux thus stems from the continuum. I conclude,
that this object is much more hydrogen under-abundant as given in
Guerrero \& Manchado (1996). The obtained continuum flux also do
not match the prediction in Koller \& Kimeswenger (2001).

As Hinkle et al. (2002), using [\ion{O}{III}] and [\ion{N}{II}]
HST images, pointed out already, V605 Aql seems to consist of a
dense inner core and a second clump separated from this region by
about 0.5\arcsec. This is much smaller than the values given by
Guerrero \& Manchado (1996), deconvolving ground based images.
Thus a completely closed special geometry can give only first
estimates. The absence of strong blue continuum straylight from
clumps illuminated by a, in line of sight hidden, central source
excludes the possibility of an open geometry. On the other hand
the strong [\ion{O}{III}]$_{4636}$ and [\ion{Ne}{III}]$_{3869}$
emission lines cannot originate from the inner edge of an
optically thick dust shell. Those strong lines were not detected
in the 1994 spectra of Guerrero \& Manchado. This is again an
indication of rapid changes in the shell.

\begin{figure}[!t]
  \includegraphics[width=\columnwidth]{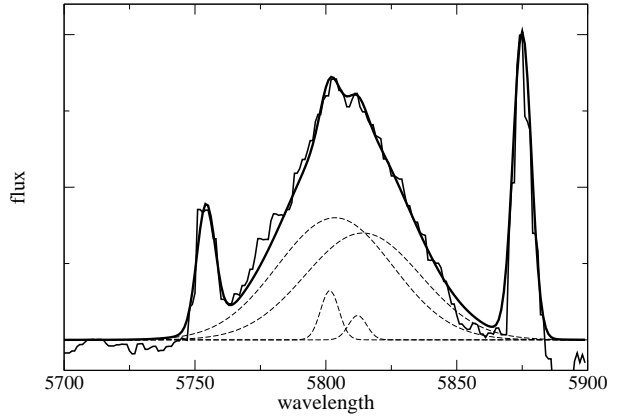}
  \caption{The carbon line \ion{C}{IV}$_{5801.5+5812.1}$ (excitation 39eV)
  is best fitted by  two components:
  The vast majority of the radiation originates from a wide
  component with a FWHM of 2600 km/s (FWZI of 6100 km/s), both lines
  also has a narrow component, best fitted by the same width as the
  neighboring lines of \ion{He}{I} and \ion{N}{II}. While the
wide main component of each line fits to the systemic velocity of
the old PN A58, the narrow component is blueshifted in the same
way as the other lines of V605 Aql. This difference in blueshift
implies a model with a optically thick dusty shell covering only a
very small outer section (Koller \& Kimeswenger 2001) and a wide
optically thin hot bubble. The complete absence of other wind
lines like \ion{O}{IV}$_{5291}$ as seen in the older twin A30
leads to the conclusion that the core of V605 Aql has been eroded
down to the pure carbon core of the star during the 1919 event. }
  \label{fig1}
\end{figure}

\end{document}